\documentclass[aps,letterpaper,floatfix,11pt]{revtex4}
\usepackage{amssymb,amsmath}
\usepackage{graphicx,floatflt}
\usepackage{citesort}

\newcommand{\unit}[1]{\,\mathrm{#1}}

\begin{document}


\title{Photoionization of strontium for trapped-ion quantum information processing}

\author{K. Vant\footnote{To whom correspondence should be
addressed. E-mail: kendra\_vant@alum.mit.edu}, J. Chiaverini, W. Lybarger, and D. J. Berkeland}

\affiliation{Applied Modern Physics, Los Alamos National Laboratory, Los Alamos, New Mexico 87545}

\begin{abstract}
We report a demonstration of simple and effective loading of strontium ions into a linear radio frequency Paul trap using photoionization.  The ionization pathway is $5s^2\ ^{1}{S}_0 \leftrightarrow 5s5p\ ^{1}{P}_1 \leftrightarrow 5p^2\ ^{1}{D}_2$, and the $5p^2\ ^{1}{D}_2$ final state is auto-ionizing.  Both transitions are driven using diode lasers: a grating-stabilized $922\unit{nm}$ diode doubled in a single pass through potassium niobate to $461\unit{nm}$ and a bare diode at $405\unit{nm}$.  Using this technique, we have reduced the background pressure during the ion loading process by a factor of 2 compared to the conventional technique of electron bombardment. Initial ion temperatures are low enough that the ions immediately form crystals.  It is also possible to observe the trapping region with a CCD camera during ion creation, allowing specific ion number loading with high probability.
\end{abstract}
\maketitle
\section{Introduction}
\label{sec:intro}
Ion traps provide one of the most promising architectures for quantum computers and quantum simulators~\cite{arda,bouw00,cira04}.  One of the requirements for trapped-ion quantum information processing is a way to reliably load isotopically pure collections of a particular ionic species to initialize a qubit register. Traditionally, ion traps have been loaded using electron bombardment of a stream of neutral atoms.  This technique results in charging of the trap and surrounding structures.  In addition, any background gas atoms and molecules are equally likely to be ionized, resulting in trapping of impurity ions.  
Photoionization of neutral atoms for the loading of ion traps has been demonstrated in a number of elements \cite{kjar00,guld01,luca04,seid06}.  For typical experimental parameters, photoionization is more effective than electron bombardment, lowering the amount of the neutral element that is plated out in the vacuum system \cite{guld01}. It also virtually eliminates charging of the trap electrodes and nearby structures \cite{kjar00,guld01}.  In this paper we describe a simple technique for photoionization of strontium, which is attractive for ion trapping experiments because all of the wavelengths required for cooling and photoionization can be produced by diode and diode-pumped solid state lasers~\cite{berk02}.
\section{Apparatus}
\label{sec:trap}
The ion trap is a linear radio-frequency (rf) Paul trap mounted inside an ultrahigh vacuum system.   Typical trap frequencies are $\omega_{radial}=2\pi\times 2\unit{MHz}$ and $\omega_{axial}=2\pi\times 400\unit{kHz}$.  Light at $422\unit{nm}$ for Doppler-cooling the ions is provided by a doubled Ti:sapphire laser, and light at $1092\unit{nm}$ for optically pumping the ions out of the metastable ${D}_{3/2}$ states comes from a Nd$^{3+}$-doped fiber laser (see Figure 1(a) for the relevant strontium ion level structure).  Neutral strontium is produced by running a current through an oven made of tantalum foil containing pieces of bulk strontium metal.  Trapped strontium ions scatter photons from the $422\unit{nm}$ cooling beam and are imaged on an intensified-charge-coupled device camera.  Details of the trap construction, vacuum system, lasers, etc.\ have been previously discussed \cite{berk02}.
\section{Ionization pathway and photoionization lasers}
\label{sec:ionization}
The ionization pathway used is a resonant two-step excitation $5s^2\ ^{1}{S}_0 \leftrightarrow 5s5p\ ^{1}{P}_1 \leftrightarrow 5p^2\ ^{1}{D}_2$ as illustrated in Figure 1(b).
\begin{figure}
\label{fig:pathway}
\includegraphics[width=5in]{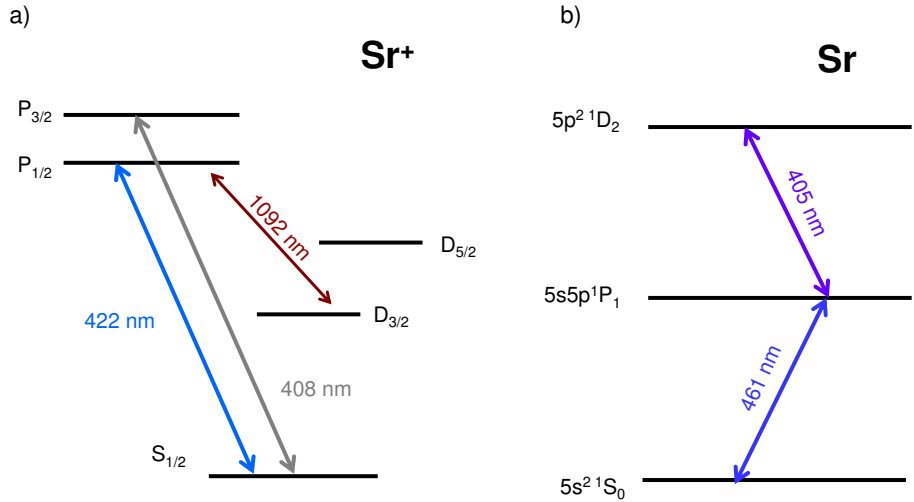}
\caption{Energy level diagram for (a) the cooling transitions used to Doppler cool $^{88}\unit{Sr}^+$ ions and (b) the two-step ionization pathway used to photoionize neutral strontium.}
\end{figure}
The $461\unit{nm}$ light used to drive the $5s^2\ ^{1}{S}_0 \leftrightarrow 5s5p\ ^{1}{P}_1$
transition is derived by frequency doubling an anti-reflection coated, grating-stabilized laser at $922\unit{nm}$ (Sacher LaserTechnik laser in setup described in \cite{arno98}).   We use non-critical phase matching in potassium niobate at approximately $160^{\circ}\unit{C}$.  A single pass through a $10\unit{mm}$ long crystal provides approximately $15\unit{\mu W}$ of light at $461\unit{nm}$ for $55\unit{mW}$ input power.  The $922\unit{nm}$ laser is passively stabilized by utilizing a compact laser structure~\cite{arno98} and a sealed metal enclosure.

The second transition, to the auto-ionizing $5p^2\ ^{1}{D}_2$ level, has a full width at half maximum of around $1\unit{nm}$~\cite{dai96} and a peak cross section of $5600\unit{Mb}$~\cite{mend95}.
Up to $15\unit{mW}$ of $405\unit{nm}$ light is produced by a bare diode (Sanyo DL-3146-151).  Temperature tuning to around $12^{\circ}\unit{C}$ is used to shift the wavelength of the diode to $405.2\unit{nm}$, the center of the $5s5p\ ^{1}{P}_1 \leftrightarrow 5p^2\ ^{1}{D}_2$ transition.  This light is coupled into a single-mode fiber to improve the spatial profile and overlapped with the $461\unit{nm}$ light on a dichroic splitter.  In this way, we can deliver up to $8\unit{\mu W}$ of $461\unit{nm}$ light and up to $3\unit{mW}$ of $405\unit{nm}$ light to the trapping region.
\section{Results and discussion}
\label{sec:results}
In order to photoionize strontium, we overlap the two photoionization beams with the $422\unit{nm}$ cooling and $1092\unit{nm}$ repumping light at the center of the trap.  The $461\unit{nm}$ spot size at the trap is $2w_0\sim140\unit{\mu m}$, giving an intensity of approximately $325\unit{W/m^2}$ for a $5\unit{\mu W}$ input power.  The saturation intensity of this transition is $I_{sat}=428\unit{W/m^2}$ using $I_{sat}=\frac{\pi}{3}\frac{hc\Gamma}{\lambda^3}$, $\Gamma=2\pi\times 32\unit{MHz}$.  The $405\unit{nm}$ spot size at the trap is $2w_0\sim70\unit{\mu m}$. Typically we have $1.5\unit{mW}$ of $405\unit{nm}$ light at the trap, giving an intensity of approximately $3.9\times 10^5 \unit{W/m^2}$.

\begin{figure}
    \centering
        \includegraphics[width=5in]{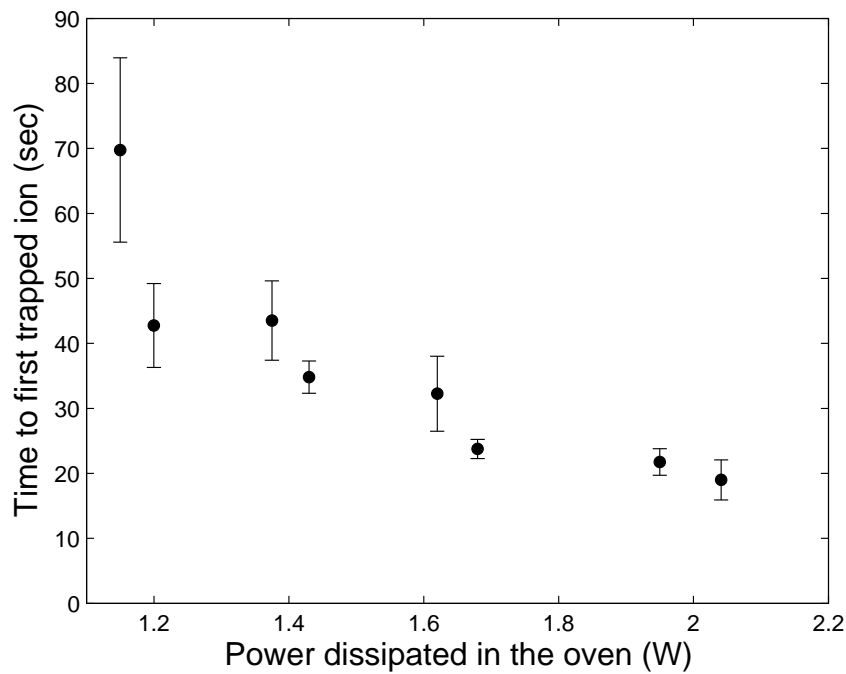}
        \caption{Average time at which the first trapped ion is observed as a function of power dissipated in the oven.}
    \label{fig:first_ion}
\end{figure}

The cooling laser is red-detuned from the $422\unit{nm}$ transition by roughly $200\unit{MHz}$ in order to Doppler-cool a significant fraction of the thermally broadened ensemble of ions. A beam of neutral strontium is produced by running a current through the strontium oven.  The photoionization light is then applied continuously and the first ions are observed $\sim20$ seconds after current is first applied to the oven. 
Trapped ion production is then very rapid---up to several ions per second are trapped and rapidly cool into a crystal structure \cite{raiz91}.  The 20 second delay is due to the finite time required for the oven to reach a sufficiently high temperature.  Figure~\ref{fig:first_ion} shows the time to the first ion being trapped as a function of power dissipated in the oven.
For comparison, electron bombardment loading was typically achieved in this system by running the oven at $\sim 2\unit{W}$ for $160$ seconds and firing the electron gun for several seconds.

When loading with an electron gun in the present setup, the imaging camera must be turned off because the light from the gun is sufficient to completely saturate the camera.   Hence loading is blind, and generally a large cloud of ions is created that must then be reduced in number and crystallized.   With photoionization, it is possible to observe ion creation and capture in real time, allowing us to load the desired number of ions with high probability.

We have made no effort to make the creation of strontium ions isotopically selective.  Due to the geometry of the experimental apparatus, the photoionization lasers enter at an angle of approximately $68^{\circ}$ to the strontium beam which is collimated by the trap electrodes to $\sim2^{\circ}$.  The three most abundant strontium isotopes are $^{86}\unit{Sr}$ (9.9\%), $^{87}\unit{Sr}$ (7.0\%), and $^{88}\unit{Sr}$ (82.6\%), so our currently desired $^{88}\unit{Sr}$ would be the predominantly loaded species in any case.   In fact we have observed that $^{88}\unit{Sr}^+$ makes up approximately 92\% of the ions loaded.  We believe this is most likely due to some isotope selectivity in the capture process.  For instance, when the cooling laser is $200\unit{MHz}$ red-detuned of the $^{88}\unit{Sr}^+$ resonance, it is still blue-detuned of the $^{86}\unit{Sr}^+$ resonance.  Hence $^{86}\unit{Sr}^+$ ions will actually be heated by the $422\unit{nm}$ photons and are trapped only because of sympathetic cooling by the $^{88}\unit{Sr}^+$ ions.  We note that the S-P transitions in $^{87}\unit{Sr}^+$ are far from resonance with our lasers and hence these ions are not heated (or directly cooled) during the loading process.

\begin{figure}
    \centering
        \includegraphics[width=5in]{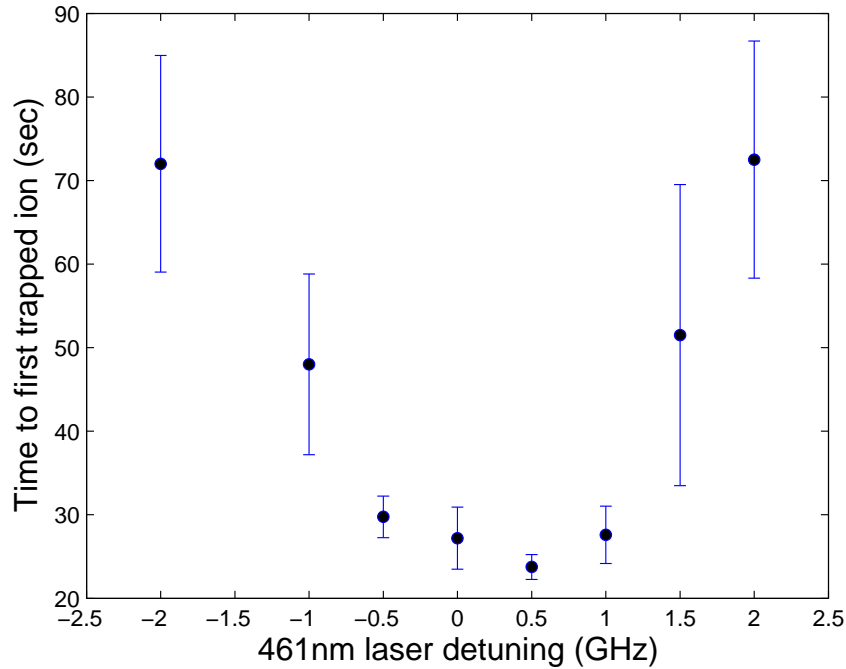}
        \caption{Time to the first trapped ion observed as a function of the frequency of the $461\unit{nm}$ laser. Zero detuning is approximately $650503.7\unit{GHz}$ ($461.181\unit{nm}$ vacuum wavelength) as measured on a Coherent WaveMaster.}
    \label{fig:461det}
\end{figure}

We have also investigated the efficiency of photoionization as a function of the $461\unit{nm}$ laser frequency (see Figure~\ref{fig:461det}).  Given the temperature of the strontium atoms, the Doppler-broadened width of the resonance should be approximately $1\unit{GHz}$.
We believe that the observed broadened and flat-bottomed line is an artifact due to the data collection technique.   During these experiments, the $422\unit{nm}$ cooling laser was not locked, leading to a slow but significant drift in frequency.  This precluded making ion number measurements from the collected scattered light intensity. Consequently, we measure the time to first trapped ion observed rather than the rate of ion formation.  (The $\sim 20$ second lower limit for ion production is due to the finite heat up time of the oven.)   Even with these limitations, the presented data gives a good indication of the level of precision required on the $461\unit{nm}$ laser frequency measurement.   We reliably locate the resonance from day to day by reading the wavelength to $0.001\unit{nm}$ on a wavemeter (Coherent WaveMaster).

We further note that we have observed the scattered cooling light from an individual trapped ion to cease intermittently when the $405\unit{nm}$ photoionization light is present \cite{toyo01}.  Although the diode laser is centered at $405.2\unit{nm}$, there is evidently enough amplified spontaneous emission (ASE) at $408\unit{nm}$ to drive the $S_{1/2}\leftrightarrow P_{3/2}$ transition (see Figure 1(a)).  Ions in the $P_{3/2}$ state decay to the $D_{5/2}$ state, which has a $395\unit{ms}$ lifetime.  In these experiments, we do not optically pump the population out of this state.  Thus ions can be shelved in $D_{5/2}$ where they remain dark to the cooling light for short periods of time.   This effect can be used to improve the alignment of the photoionization beams with the trap minimum.
\section{Conclusion}
We have demonstrated a simple and robust method for photoionization loading of an rf Paul trap with strontium ions.  Both of the lasers used are low-maintenance, stable diode lasers which are readily available and easy to operate. Single-pass frequency doubling enhances the reliability and simplicity of the system.  Observation of the trapping region during ion creation is possible, which facilitates loading a given ion number with high probability.  We note that photoionization loading results in a smaller increase in the background pressure than electron bombardment loading.  This fact, coupled with reduced charging of the trap and surrounding structures~\cite{guld01} allows ions to be loaded at lower temperatures than in electron bombardment.  This feature makes photoionization critical for loading the miniaturized and shallow microfabricated traps which look likely to be the future of scalable ion-trap quantum computing and quantum simulation.
\section{Acknowledgements}
It is a pleasure to thank Malcolm Boshier for critically reading the manuscript.
\bibliographystyle{unsrt}
\bibliography{kbib}
\end{document}